\documentclass[aps,twocolumn, floats,preprintnumbers,showpacs,prl]{revtex4}
\usepackage{graphicx}
\usepackage{url}
\usepackage{amsmath}
\usepackage{amsfonts}
\usepackage{amssymb}
\def\beq{\begin{equation}}
\def\eeq{\end{equation}}
\def\beqa{\begin{eqnarray}}
\def\eeqa{\end{eqnarray}}

\def\ltap{\ \raise.3ex\hbox{$<$\kern-.75em\lower1ex\hbox{$\sim$}}\ }
\def\gtap{\ \raise.3ex\hbox{$>$\kern-.75em\lower1ex\hbox{$\sim$}}\ }

\begin{document}
\preprint{FERMILAB-PUB-10-003-T}

\title{Topological Pions}

\author{Yang Bai and Adam Martin
\\
\vspace{2mm}
Theoretical Physics Department, Fermilab, Batavia, Illinois 60510
}



\begin{abstract}
We study the collider signatures of new pions, composite particles which emerge from a TeV-scale, confining gauge theory with vector-like matter. Similar to the neutral pion in QCD, these new pions mainly decay into a pair of standard model (SM) gauge bosons via triangular anomaly diagrams. One of the new pions, which decays to a gluon plus a photon, has excellent discovery potential at the LHC.
\end{abstract}
\maketitle

{\it{\textbf{Introduction.}}}
If new strong dynamics lies at the TeV-scale, its discovery may repeat the same history as the discovery of QCD. New pions, the Nambu-Goldstone Bosons (NGBs) from spontaneous chiral symmetry breaking in the new sector, would be the lightest composite particles and the first encountered at colliders. If the underlying strong-sector fermions have vector-like quantum numbers under the new gauge group and under all SM gauge groups, the resulting pions can only decay via triangular anomaly diagrams into pairs of SM gauge bosons. This is analogous to the $\pi^0$ in QCD, which decays predominantly to two photons, $\pi^0\rightarrow \gamma\gamma$~\cite{Adler:1969gk}. The discovery these di-boson resonances would provide us with the first hint of strong dynamics beyond the standard model. 

Other strongly coupled extensions of the SM, such as technicolor also predict various new pions (there called technipions).  Technipions, however, typically have sizable couplings to SM fermions~\cite{Hill:2002ap}, and the resulting fermionic decay modes completely dominate over any anomaly induced channels. In technicolor,  SM fermions must communicate with technifermions in order to become massive, thus direct interactions between technipions and SM fermions are inevitable. However, if new pions come from a sector unrelated to electroweak symmetry breaking, as in Ref.~\cite{Kilic:2008pm},  no direct interaction involving SM fermions is necessary. In this case, only SM gauge bosons directly talk to the new strongly interacting sector. 

In this letter, we use a simple and representative example to study new pions from a new asymptotic-free gauge theory $SU(N_n)$. Due to their NGB nature, the new pions are light degrees of freedom below the confinement scale $\Lambda_n$ and are the most relevant new states at colliders~\footnote{Other heavier resonances like $\rho$ mesons can also be studied if the $\Lambda_n$ is low enough to be accessible at colliders~\cite{Kilic:2008pm}.}. For underlying fermions with vector-like charges under all SM gauge groups, the new pions are in the adjoint representations. A particularly interesting pion multiplet, and the focus of this note, lies in the bi-adjoint representation of $SU(3)_c\times SU(2)_W$. We denote this pion as $\Pi^{A, \pm}$ and $\Pi^{A, 0}$, with ``$A$" indicating the color index.  Pions in this representation have a large production cross section at hadron colliders, and their anomaly-induced decays, $\Pi^{A, \pm}\rightarrow j+W^\pm$ and $\Pi^{A, 0} \rightarrow j + \gamma, Z$, {\em always} include an electroweak gauge boson. This combination of quantum numbers leads to unique, relatively clean resonance signals. 

As only the SM gauge bosons can directly interact with the new pions, pair production of the new pions will be dominant at colliders.  Jet plus weak gauge boson resonances have been studied before in context of excited quarks~\cite{Baur:1987ga, Baur:1989kv}. However, because of different underlying dynamics, excited quarks can be singly produced. Hence, excited quarks have totally different final states and experimental constraints from the new pions explored in this paper. 

Electroweak singlet and doublet color octets (pseudo) scalars have been considered in the past~\cite{Kilic:2008pm,Manohar:2006ga, Dobrescu:2007yp, Gerbush:2007fe,Zerwekh:2008mn}. These states share the same production mechanism ($gg \rightarrow \Pi^A \Pi^A$) as the bi-adjoint pions considered here, but their decays are quite different. Due to the electroweak triplet nature of the new pions we consider, electroweak gauge bosons will be present in every decay.  Meanwhile, electoweak singlet color octets decay predominantly to gluons, with decays to electroweak gauge bosons suppressed by $\sim \alpha_{em}/{\alpha_s}$, and electroweak doublet octets can mainly decay to pairs of SM fermions because renormalizable operators are allowed.

The paper is organized as followings. We first analyze the spectrum and topological decay modes of a simple model with a new pion in the bi-adjoint representation of $SU(3)_c\times SU(2)_W$. Afterwards, we investigate the current experimental bounds on $\Pi^{A, 0}$ and perform a collider study to estimate its discovery potential at  the Tevatron and the LHC. The LHC studies are performed with both 7 TeV and 14 TeV center of mass energy. 

{\it{\textbf{A model with a new strong dynamics.}}}
Guided by QCD, we assume there exists a new asymptotically-free gauge group, $SU(N_n)$, with $N_n \geq 2$. We introduce some fermions $\psi_{L,R}$, which are in the fundamental representation of $SU(N_n)$ and, unlike in the SM, the new fermions are assumed to have vector-like representations under all gauge symmetries. Hence, all gauge anomalies are cancelled. The gauge groups and field content are listed in Table~\ref{tab:fieldcontent}.
\begin{table}[htdp]
\renewcommand{\arraystretch}{1.5}
\begin{center}
\begin{tabular}{ccccc}
\hline \hline
   &   $SU(N_n)$    &   $SU(3)_c$    &   $SU(2)_W$    & $U(1)_Y$    \\  \hline
$\psi _{L, R}$    &  $N$          &  3                 &    2                  & $-1/2$      \\ \hline   \hline
\end{tabular}
\end{center}
\caption{Field content of the model with a new strong dynamics.   \label{tab:fieldcontent}}
\end{table}%
The hypercharge of $\psi_{L,R}$ is chosen to be the same as the left-handed leptons, though it is not essential for our following analysis. In order for the $SU(N_n)$ gauge group to be asymptotically-free, the number of fundamental flavors, $N_f$, should be less than $11\,N_n/2$. This is indeed the case for the model in Table~\ref{tab:fieldcontent}, where $N_f =6$. Similarly, to maintain the asymptotic freedom of $SU(3)_c$, we require $2\leq N_n \leq 5$.

Analogous to $\Lambda_{\rm QCD}$ in QCD, there exists a scale $\Lambda_n$, where  the $SU(N_n)$ gauge coupling becomes strong, inducing confinement and spontaneous chiral symmetry breaking. Motivated purely by phenomenological interests,  we take the confinement scale, $\Lambda_n$, to be around the TeV scale. The bi-fermion condensate is approximately related to the scale $\Lambda_{n}$ as $\langle \bar{\psi}^i_{L}\,\psi^i_{R} \rangle \sim \Lambda_n^3$ where $i = 1, 2, \cdots, N_f$ denote the flavor indexes. Unlike what happens in technicolor, the fermion condensation in this model does not break any SM gauge symmetries. This is a consequence of our choice of vector-like representations in Table~\ref{tab:fieldcontent}.

In the limit of vanishing SM gauge couplings, the global symmetry of this model is $SU(N_f)_L \times SU(N_f)_R \times U(1)_A \times U(1)_{B_n}$. The $U(1)_A$ is broken by the instanton effects and will be discarded in the following discussions. The baryon number symmetry $U(1)_{B_n}$ associated with those new fermions is unbroken by the condensation and stays as a good symmetry. The global symmetry $SU(N_f)_L \times SU(N_f)_R$ is spontaneously broken to the diagonal group $SU(N_f)_V$, resulting in $N_f^2 - 1 = 35$ NGB's, \textit{new pions}, in the low energy theory. Decomposing the 35 pions into representations of SM gauge groups $SU(3)_c \times SU(2)_W \times U(1)_Y$, we have 
\beqa
35 &\rightarrow& (8, 1)_0 \,+\, (1, 3)_0 \,+\, (8, 3)_0  \,.
\eeqa
We parameterize these new pions as $\Pi^A$, $\Pi^a$ and $\Pi^{A, a}$ with $A = 1, \cdots , 8$ and $a = 1, 2, 3$. In the following discussions, we will concentrate on phenomenology of the neutral part of the $(8, 3)_0$ pion, $\Pi^{A, 0}$. Other pions can be studied in a similar manner.  

Turning on SM gauge couplings, all pions become massive because the global symmetry is explicitly broken by the gauge couplings. The masses of new pions are generated at one loop level from gauge boson loops. For the $(8, 3)$ pion, its mass is estimated to be 
%
%
$M_{\Pi^{A, 0}} \,=\, c\, \sqrt{3\, g_s^2\,+\, 2\, g_2^2}\,f_\Pi$ with $c$ order of unit and $g_s$ and $g_2$  are gauge couplings of $SU(3)_c$ and $SU(2)_W$. The pion decay constant $f_\Pi$ is related to the confinement scale $\Lambda_n$ as $\Lambda_n \sim 4\pi f_\Pi$. Because the SM gauge couplings are small at high energies, the new pions all have masses well below the confinement scale. The parameter $c$ depends on UV physics and may be different  for each pion multiplet. This allows us to study each pion multiplet separately. Within the $(8,3)_0$ multiplet, the mass of the charged component is enhanced compared to the mass of the neutral component by $\mathcal O$(100~MeV) due to electromagnetic radiative corrections.

{\it{\textbf{Topological decays.}}}
In the SM, the main decay channel of $\pi^0$, $\pi^0 \rightarrow \gamma\gamma$, is mediated by triangular anomalies. In our model the leading interactions among new pions and the SM particles (besides the pion kinetic terms) also come from topological interactions via triangular anomaly diagrams. The topological interactions are also the only interactions which violate pion number. The general form for topological interactions among $\Pi$s and standard model gauge bosons can be written as
\beq
{\cal L}_{\rm topo}\,=\, -\,\frac{g_B\,g_C\,\Pi_A}{16\,\pi^2\,f_\Pi}\,\epsilon_{\mu\nu\rho\sigma}\,F_B^{\mu\nu}\,F^{\rho\sigma}_C\,{\rm Tr}\{ t^A\,t^B\,t^C\} \,,
\label{eq:generalformula}
\eeq
where $\Pi_A$ represents different kinds of pions and $F_B^{\mu\nu}$ represents SM gauge bosons; $g_{B, C}$ are gauge couplings, while $t^A$ and $t^{B, C}$ are fermion representations under the axial global current and SM gauge symmetries, respectively. The ``Tr" acts on every group matrix and also takes into account the ``color" multiplication factor $N_n$ for the loop calculations. Applying the general formula in Eq.~(\ref{eq:generalformula}) for $\Pi^{A, a}$, we arrive at the following interactions:
\beqa
{\cal L}_{\rm topo} &=&\,- \frac{\sqrt{2}\,g_s\,g_2\,N_n}{4}\,\frac{\Pi^{A, a}}{16\,\pi^2\,f_\Pi}\,\epsilon_{\mu\nu\rho\sigma}\,G^{\mu\nu}_A\,W^{\rho\sigma}_a\,.
\eeqa
Here, $G$ and $W$ represent the $SU(3)_c$ and $SU(2)_W$ gauge fields. Due to gauge symmetries, this is the only operator at the dimension-5 level. Higher-dimensional operators coming from the chiral Lagrangian have negligible effects on the width of new pions and will be neglected. 

\begin{figure}[t!]
\vspace{0.0cm}
\centerline{ \hspace*{0.0cm}
\includegraphics[width=0.45\textwidth]{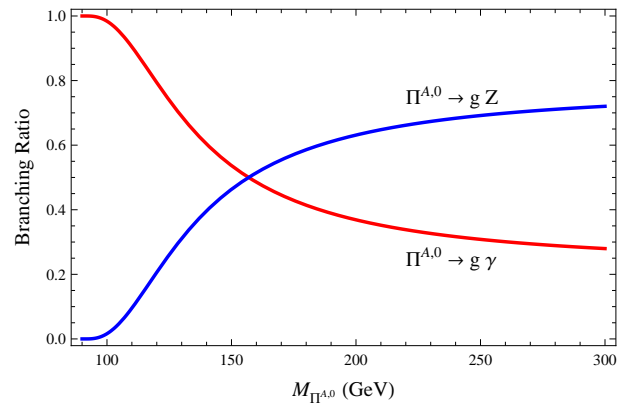}
}
\caption{The branching ratios of $\Pi^{A, 0}$. 
\label{fig:PBranchingRatio}
}
\end{figure}

After electroweak symmetry breaking, the interactions involving $\Pi^{A, 0}$ can be rewritten in terms of the physical $Z$ bosons and the photon $A$:
\beqa
\frac{\sqrt{2}\,g_s\,g_2\,N_n}{4}\,\frac{\Pi^{A, 0}}{16\,\pi^2\,f_\Pi} \epsilon_{\mu\nu\rho\sigma}\,G^{\mu\nu}_A \,(c_W\,Z^{\rho\sigma}-s_W\,A^{\rho\sigma})  \,,
\eeqa
where $c_W (s_W) \equiv \cos{\theta_W} (\sin{\theta_W})$, with $\theta_W$ as the Weinberg angle. From the  interactions of new pions with SM gauge bosons, we calculate the partial decay widths 
\beqa
\Gamma(\Pi^{A, 0} \rightarrow g + Z)&=&\frac{N_n^2\alpha_s\alpha_2 c_W^2}{256 \pi^3}\frac{M^3_{\Pi^{A, 0}}}{f^2_\Pi}\left(1-\frac{M_Z^2}{M^2_{\Pi^{A, 0}}}\right)^3\,, \nonumber \\
\Gamma(\Pi^{A, 0} \rightarrow g + \gamma)&=&\frac{N_n^2\alpha_s\alpha_2 s_W^2}{256 \pi^3}\frac{M^3_{\Pi^{A, 0}}}{f^2_\Pi}   \,.
\eeqa
The branching ratios as a function of pion mass are shown in Fig.~\ref{fig:PBranchingRatio}. For pion masses far above the $M_Z$, the branching ratios become constant, ${\rm Br}(\Pi^{A, 0} \rightarrow g + \gamma ) \approx 22.5\%$ and ${\rm Br}(\Pi^{A, 0} \rightarrow g + Z ) \approx 77.5\%$. The total width of $\Pi^{A,0}$ is always small -- $\mathcal O(10)$ MeV for a mass of 1 TeV. 

\begin{figure}[t!]
\vspace{0.0cm}
\centerline{ \hspace*{0.0cm}
\includegraphics[width=0.45\textwidth]{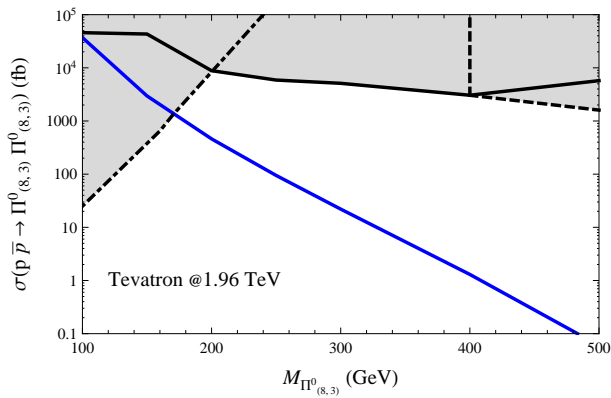}
}
\caption{Production cross section of a pair of $\Pi^{A, 0}$ at Tevatron with 1.96~TeV center of mass energy. The shaded region is excluded by various resonance searches at Tevatron as explained in the text. 
\label{fig:productTevatron}
}
\end{figure}

{\it{\textbf{Experimental constraints.}}}
%
%
Using MADGRAPH/MADEVENT~\cite{Alwall:2007st}, the total production cross section (including quarks as partons) of this particle at the Tevatron with 1.96~TeV  is  shown in Fig.~\ref{fig:productTevatron}. Areas excluded by direct Tevatron searches are also shown.
 The first bound on $\Pi^{A,0}$ production, indicated by the solid black line in Fig.~\ref{fig:productTevatron}, comes from the search for excited quarks decaying as $q^* \rightarrow q \gamma$~\cite{CDFcons} performed during Run-1. The fact that Ref.~\cite{CDFcons} investigated single production has been taken into account. Limits from a second excited quark channel, $q^*\rightarrow q Z$ with $Z\rightarrow e^+ e^-$~\cite{D0cons} have also been included and are shown in the dashed black line. The final constraint, the dot-dashed line, comes from di-photon resonance searches at D0~\cite{Abazov:2009kq}. To obtain this constraint, we first impose the same cuts in \cite{Abazov:2009kq} and form the di-photon invariant mass distribution. We then require the number of signal events to be below the experimental error shown in the last bin of Fig.~1 in~\cite{Abazov:2009kq}. As seen from Fig.~\ref{fig:productTevatron}, this new pion $\Pi^{A, 0}$ is constrained to have a mass above $\sim$170 GeV. The charged partner, $\Pi^{A, \pm}$, is nearly degenerate with $\Pi^{A, 0}$, but has a weaker constraint from the long-lived charged massive particle searches. 
 
{\it{\textbf{Discovery potential at the Tevatron and the LHC.}}}
In the following, we concentrate on the channel of $\Pi^{A, 0}\rightarrow g + \gamma$, although the $g+Z$ channel may also be interesting. For pair-produced $\Pi^{A, 0}$, the final state is therefore $2\,j + 2 \,\gamma$. To study this channel, we first use MADGRAPH/MADEVENT~\cite{Alwall:2007st} to generate pairs of  $\Pi^{A, 0}$. The $\Pi^{A, 0}$  are then decayed using BRIDGE~\cite{Meade:2007js} into a gluon plus a photon.
The dominant irreducible SM background comes from inclusive $n\,j + 2\,\gamma$ with $n=1,\cdots, 4$, and is generated with ALPGEN~\cite{Mangano:2002ea} . The signal and background parton-level events are further processed with PYTHIA~\cite{Sjostrand:2006za} for showering/hadronization and PGS~\cite{pgs} for detector simulation. 

At the Tevatron with 1.96 TeV, we choose a 180 GeV mass for $\Pi^{A, 0}$ as an example to illustrate the event reconstruction.
\begin{figure}[t!]
\vspace{0.0cm}
\centerline{ \hspace*{0.0cm}
\includegraphics[width=0.45\textwidth]{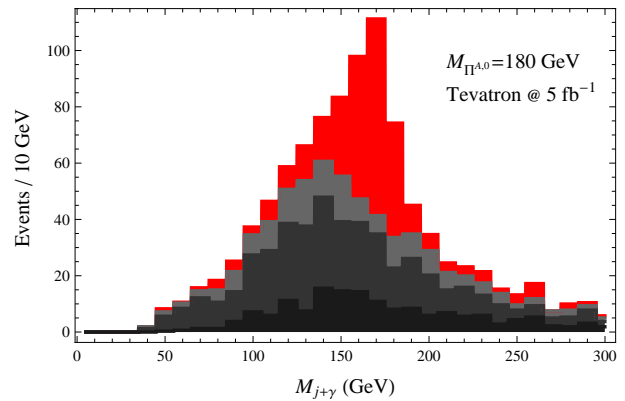}
}
\caption{Reconstructed signal and background events as a function of one jet and one photon invariant masses. The signals are in the red region and the backgrounds are in the gray and black regions. From bottom to top, the backgrounds are $1j+2\gamma$, $2j+2\gamma$ and $3j+2\gamma$. The width of this pion, $\mathcal O$(1~MeV), is completely negligible compared to energy resolution effects. 
\label{fig:P180GeV}
}
\end{figure}
To separate the signal from the background, we impose the following cuts: the two leading jets and two leading photons must all have $p_T$ above 30 GeV, the $H_T$ of these four objects must be above 180~GeV,  and the rapidities of the two photons are restricted to $|\eta_{\gamma}| \le 2.0$.  For these cuts, the signal acceptance efficiency is around 20\%. To reconstruct the $\Pi^{A, 0}$, each of the leading jets is paired with a photon. Of the two possible pairing assignments, we choose the combination which yields closer invariant masses for the two reconstructed $\Pi^{A,0}$. We record the larger jet + photon invariant mass in Fig.~\ref{fig:P180GeV}; this is justified by the reduction to the invariant mass from initial and final state radiation.

In Fig.~\ref{fig:P180GeV} and from bottom to top, we have the inclusive $1j+2\gamma$, $2j+2\gamma$ and $3j+2\gamma$ backgrounds. The $n\ge4 j + 2\gamma$ background is negligible and is not shown. Another background is $nj + 1\gamma$ with one or more jets faking a photon. This background depends sensitively on fake rates and is not shown in this plot, but it will be included in the discovery potential estimation later. The $t\bar{t}$ background has also been simulated and is found to be negligible. 

Counting events in the five bins near the peak  (3 bins below and 1 bin above) from 135 GeV to 185 GeV, we find $S/\sqrt{B} \approx 13$. A new pion with this mass could certainly be found using the current data at the Tevatron, though the actual significance will depend on the fake-photon background. Exploring a wider range of $M_{\Pi^{A, 0}}$, the required Tevatron $S/\sqrt{B} = 5~\sigma$ discovery luminosity is shown in Fig.~\ref{fig:discoverypotential}.

\begin{figure}[t!]
\vspace{0.0cm}
\centerline{ \hspace*{0.0cm}
\includegraphics[width=0.45\textwidth]{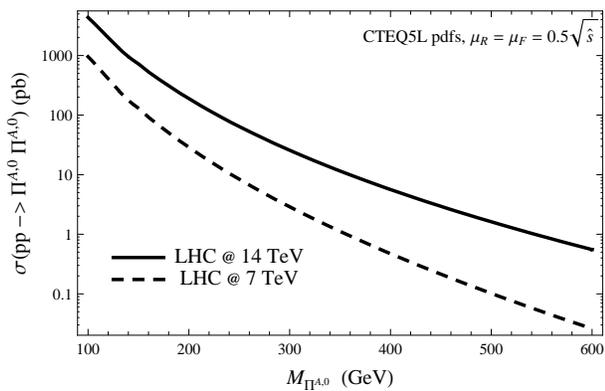}
}
\caption{Leading order production cross section at the LHC for two different center of mass energies. 
\label{fig:productLHC}
}
\end{figure}

\begin{figure}[ht!]
\vspace{0.0cm}
\centerline{ \hspace*{0.0cm}
\includegraphics[width=0.45\textwidth]{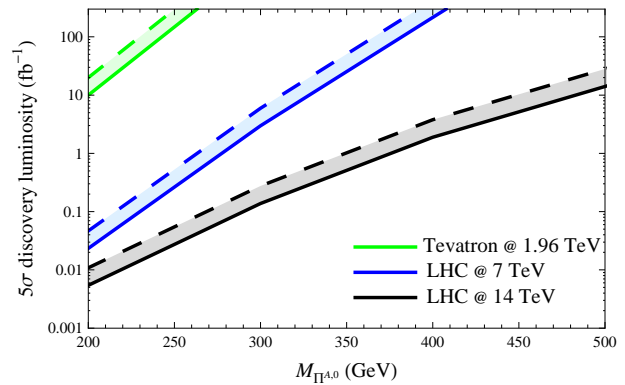}
}
\caption{Luminosity required for $5\,\sigma$ discovery as a function of $M_{\Pi^{A,0}}$ for Tevatron at 1.96 TeV, LHC at 7 TeV and 14 TeV.  
The dashed lines are the discovery potentials by including maximum jet-fake backgrounds.
\label{fig:discoverypotential}
}
\end{figure}

Turning to the LHC, the leading order production cross section is shown in Fig.~\ref{fig:productLHC} for both $7$~TeV and $14$~TeV center of mass energy.
As the only irreducible background is still $2\gamma + $ jets, we 
repeat a similar analysis as for the Tevatron.  To simplify our presentation, for a given $M_{\Pi^{A,0}}$ we apply the following mass-dependent cuts: leading photon and jet $p_{T} > M_{\Pi}/3$,  subleading photon and jet  $p_{T} > M_{\Pi}/6$ and  $H_T > \max(M_{\Pi}, 3M_{\Pi} - 600\ \text{GeV})$. After cuts, the reconstructed jet-photon invariant masses are binned in $20\ \text{GeV}$, and we estimate the significance by counting signal and background events within the five bins closest to the peak.  The luminosity required for $5\,\sigma$ discovery for different pions masses has been calculated and is shown in Fig.~\ref{fig:discoverypotential}. To account for possible background from jets faking photons, we repeat the same procedure using twice the background and show the result in the dashed lines of Fig.~\ref{fig:discoverypotential}. This corresponds to a fake background equal to the simulated two photon background and is a rough estimate of the maximum pollution from jet-fakes~\cite{fakejet}. The region between the zero-fake and maximum-fake contours, shaded in Fig.~\ref{fig:discoverypotential}, is our estimated discovery luminosity.
%


{\it{\textbf{Discussions and conclusions.}}}
As we know from QCD, topological interactions are important not only to explain a single physical observable, but also to help us understand the details of the underlying dynamics. If new gluon + $\gamma/Z/W^{\pm}$ resonances are found at the Tevatron or the LHC, the next step would be to measure the spin information. Spin information is necessary to distinguish pseudo-NGB resonances from other cases, such as the spin-1 resonances studied in~\cite{Hill:2007zv, Keung:2008ve, Lane:2009ct, Bai:2009ij}. Once the pseudo-NGB nature has been confirmed, we can measure its decay width (though challenging) to determine the ``color" $N_n$ of the new strong dynamics. 

In conclusion, we have studied the properties of  \textit{new pions}, composite particles emergent from chiral symmetry breaking in a sector of new, vector-like fermions. We focused on the new pion multiplet which lies in a bi-adjoint representation of the SM color and weak gauge groups. The electrically neutral component of this multiplet has  a striking, yet unstudied, collider signature from its decay to a gluon plus a photon. 
For the Tevatron with 10 fb$^{-1}$, we find this particle can be discovered up to a mass of $M_{\Pi^{A,0}} \sim 200$ GeV. At the LHC, we find the discovery potential to be $M_{\Pi^{A,0}} \sim 280$ GeV and $M_{\Pi^{A,0}} \sim 480$ GeV for 1 fb$^{-1}$ at 7 TeV and 10 fb$^{-1}$ at 14 TeV, respectively.

\acknowledgments 
\vspace*{0.1in}
The authors are grateful to Gustavo Burdman and Chris Hill for useful discussions.  Fermilab is operated by Fermi  Research Alliance, LLC under contract no. DE-AC02-07CH11359 with the United States Department of Energy.  

 
\vfil \end{document}